\documentclass[aps,prb,twocolumn,superscriptaddress,showpacs]{revtex4}
\pdfoutput=1
\usepackage{amsmath}
\usepackage{amssymb}
\usepackage{hyperref}
\usepackage{url}
\usepackage{graphicx}
\usepackage{dcolumn} 
\usepackage{bm}      
\usepackage{multirow}
\usepackage{epstopdf}
\usepackage{color}
\usepackage{gensymb}
\usepackage{textcomp}

\begin{document}

\title{Magnetic ground states and magnetodielectric effect of $R$Cr(BO$_3$)$_2$ ($R$ = Y and Ho)}

\author{R.~Sinclair}
\affiliation{Department of Physics and Astronomy, University of Tennessee,
Knoxville, Tennessee 37996-1200, USA}

\author{H.~D.~Zhou}
\affiliation{Department of Physics and Astronomy, University of Tennessee,
Knoxville, Tennessee 37996-1200, USA}
\affiliation{National High Magnetic Field Laboratory, Florida State University, Tallahassee, Florida 32310, USA}

\author{M.~Lee}
\affiliation{Department of Physics, Florida State University, Tallahassee, Florida 32306, USA}
\affiliation{National High Magnetic Field Laboratory, Florida State University, Tallahassee, Florida 32310, USA}

\author{E.~S.~Choi}
\affiliation{National High Magnetic
Field Laboratory, Florida State University, Tallahassee, Florida 32310, USA}

\author{G.~Li}
\affiliation{School of Physics and Materials Science, Anhui University, Hefei, Anhui, 230601, People's Republic of China}

\author{T.~Hong}
\affiliation{Quantum Condensed Matter Division, Oak Ridge National Laboratory, Oak Ridge, Tennessee 37381, USA}

\author{S.~Calder}
\affiliation{Quantum Condensed Matter Division, Oak Ridge National Laboratory, Oak Ridge, Tennessee 37381, USA}

\date{\today}

\begin{abstract}

The layered perovskites $R$Cr(BO$_3$)$_2$ ($R$ = Y and Ho) with magnetic triangular lattices were studied by performing ac/dc susceptibility, specific heat, elastic and inelastic neutron scattering, and dielectric constant measurements. The results show (i) both samples' Cr$^{3+}$ spins order in a canted antiferromagnetic structure with $T_N$ around 8-9 K, while the Ho$^{3+}$ ions do not order down to $T$ = 1.5 K in HoCr(BO$_3$)$_2$; (ii) when a critical magnetic field H$_{C}$ around 2-3 T is applied below $T_{N}$, the Cr$^{3+}$ spins in the Y-compound and both the Cr$^{3+}$ and Ho$^{3+}$ spins in the Ho-compound order in a ferromagnetic state; (iii)  both samples exhibit dielectric constant anomalies around the transition temperature and
critical field, but the Ho-compound displays a much stronger magnetodielectric response. We speculate that this is due to the  magnetostriction which depends on both of the Cr$^{3+}$ and the Ho$^{3+}$ ions' ordering in the Ho-compound. Moreover, by using linear spin wave theory to simulate the inelastic neutron scattering data, we estimated the Y-compound's intralayer and interlayer exchange strengths as ferromagnetic J$_{1}$ = -0.12 meV and antiferromagnetic J$_{2}$ = 0.014 meV, respectively. The competition between different kinds of superexchange interactions results in the ferromagnetic intralayer interaction. 
\end{abstract}


\maketitle

\begin{figure}[tbp]
\linespread{1}
\par
\begin{center}
\includegraphics[width=\columnwidth]{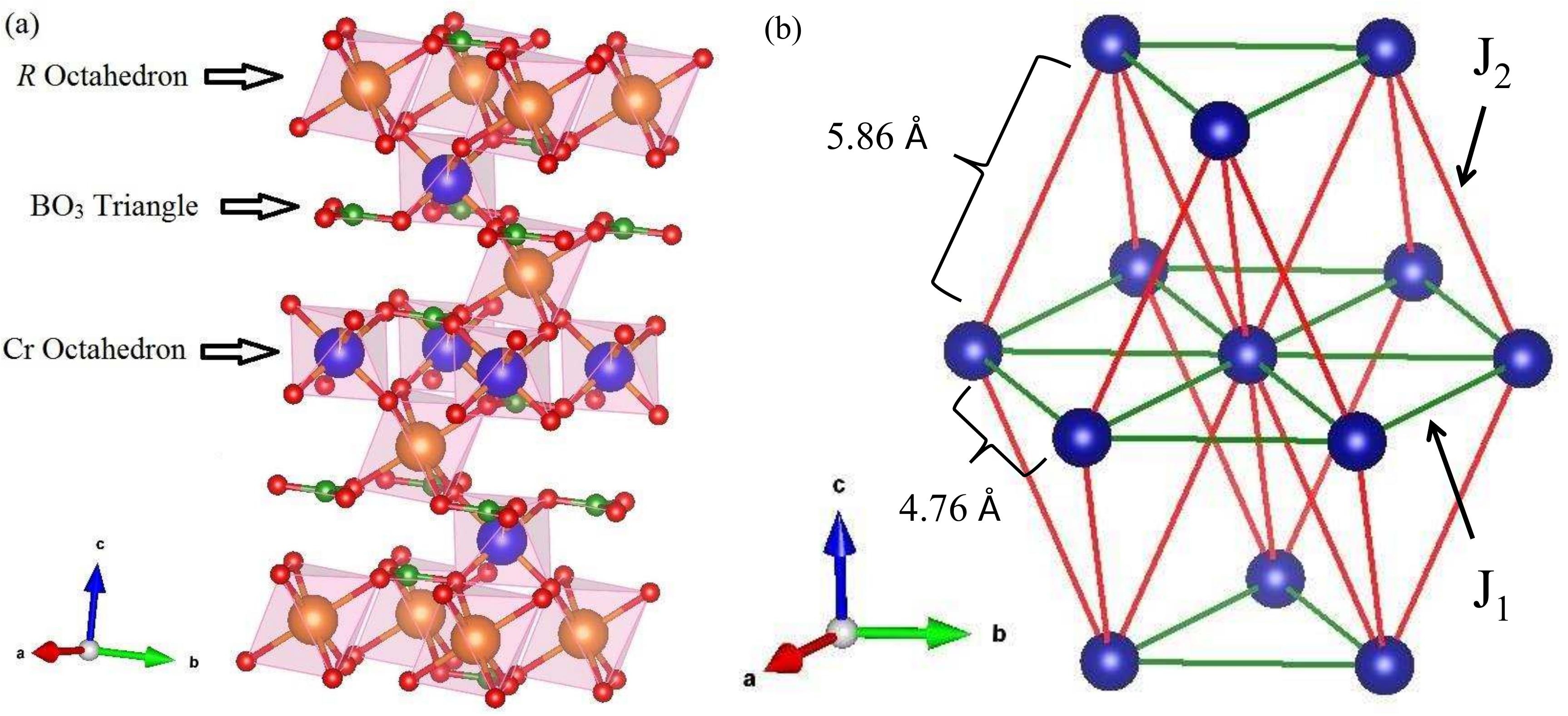}
\end{center}
\par
\caption{(color online) (a) The dolomite-type crystal structure of $R$Cr(BO$_3$)$_2$. The orange/blue $R$/Cr octahedra form a triangular lattice in the $ab$ plane as shown in (b). Only the Cr ions are plotted here. The intralayer and interlayer exchange interactions are labeled as J$_{1}$ and J$_{2}$, respectively.}
\end{figure} 

\begin{figure}[tbp]
\linespread{1}
\par
\begin{center}
\includegraphics[width=\columnwidth]{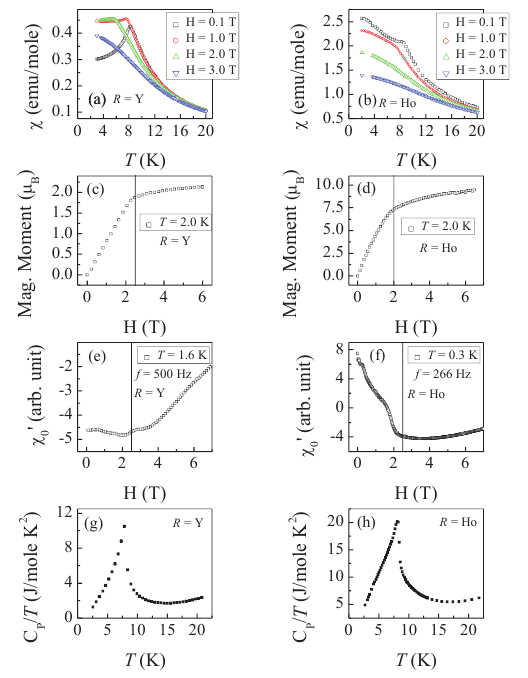}
\end{center}
\par
\caption{(color online) The temperature dependence of the dc susceptibility for (a) YCr(BO$_3$)$_2$ and (b) HoCr(BO$_3$)$_2$; the magnetization curve for (c) YCr(BO$_3$)$_2$ and (d) HoCr(BO$_3$)$_2$; the field dependence of the ac susceptibility for (e) YCr(BO$_3$)$_2$ and (f) HoCr(BO$_3$)$_2$; the specific heat capacity measured at H = 0 T for (g) YCr(BO$_3$)$_2$ and (h) HoCr(BO$_3$)$_2$.}
\end{figure}

\begin{figure}[tbp]
\linespread{1}
\par
\begin{center}
\includegraphics[width=\columnwidth]{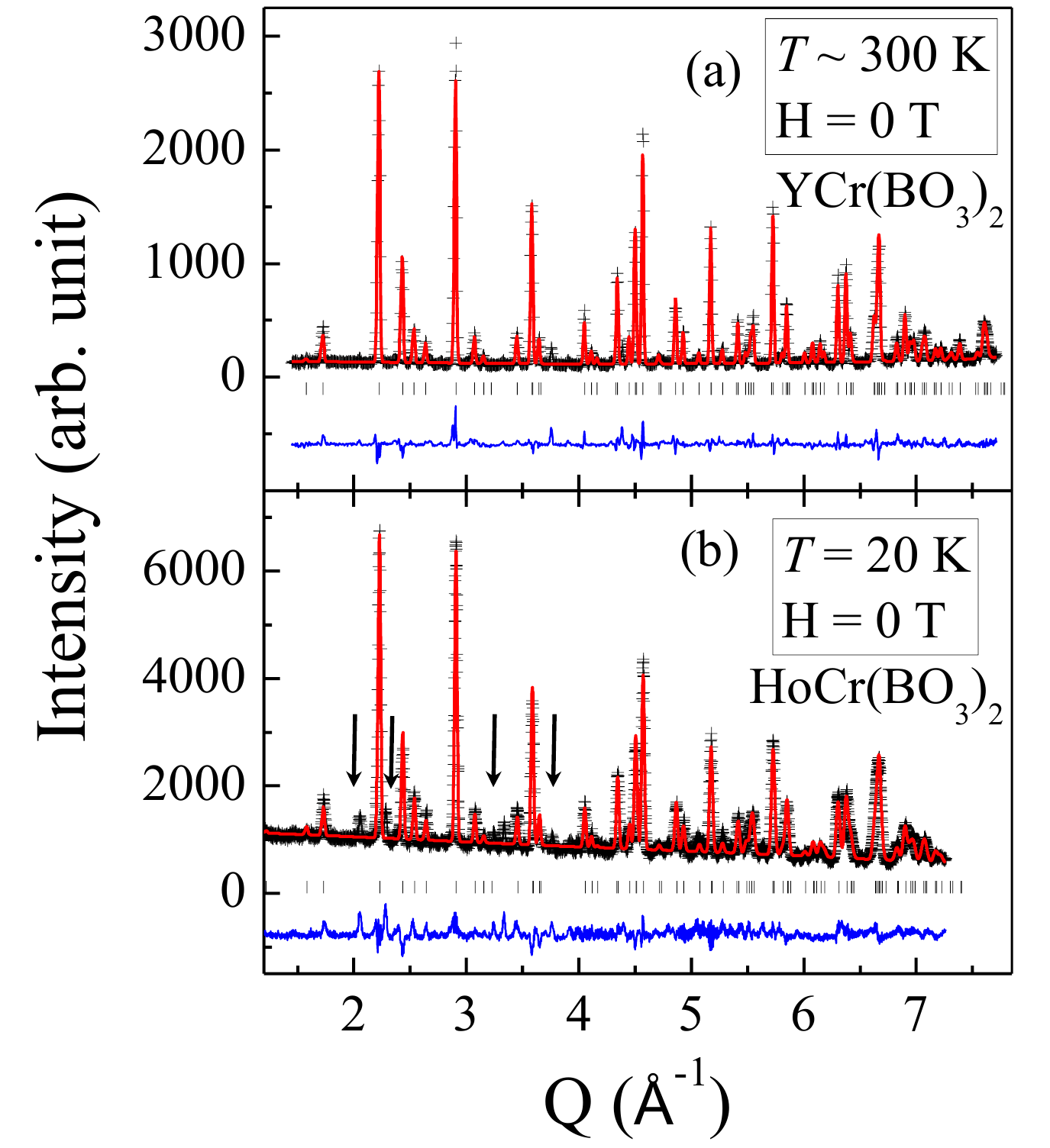}
\end{center}
\par
\caption{(color online) The elastic neutron diffraction patterns (crosses) for (a) polycrystalline 
YCr(BO$_3$)$_2$ at room temperature and zero field and for (b) polycrystalline HoCr(BO$_3$)$_2$ at $T$ = 20 K and zero field using a wavelength of 1.5405 {\AA}. The solid curves are the best fits from the Rietveld refinements using \textit{FullProf Suite}. The vertical marks indicate the position of Bragg reflections, and the bottom curves show the difference between the observed and calculated intensities. The arrows denote HoBO$_3$ impurity peaks.}
\end{figure}

\begin{figure}[tbp]
\linespread{1}
\par
\begin{center}
\includegraphics[width=\columnwidth]{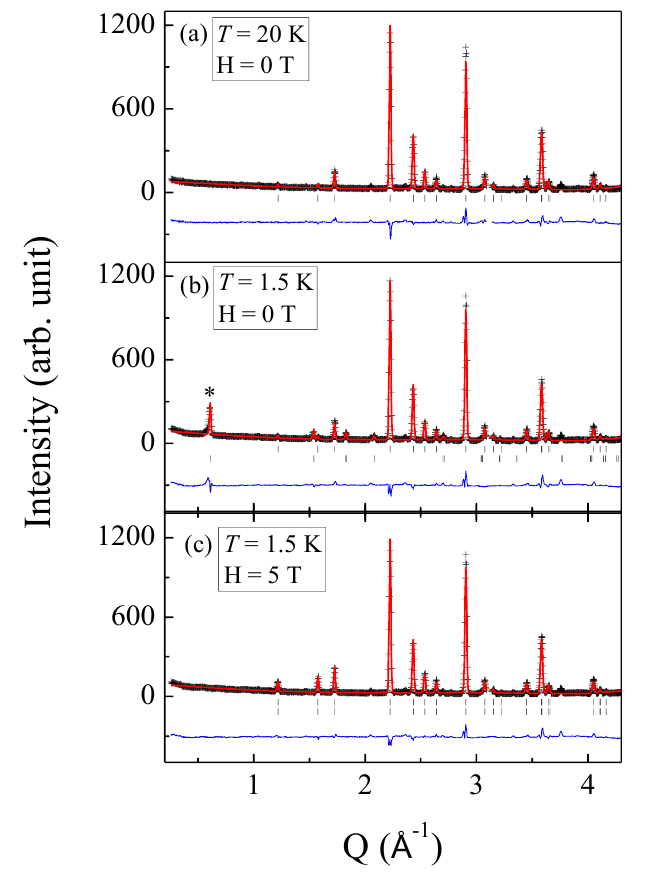}
\end{center}
\par
\caption{(color online) The neutron diffraction patterns for polycrystalline 
YCr(BO$_3$)$_2$ (crosses) at (a) $T$ = 20 K and H = 0 T, (b) $T$ = 1.5 K and H = 0 T,
and (c) $T$ = 1.5 K and H = 5.0 T using a neutron wavelength of 2.413 {\AA}. The solid curves are the best fits from the Rietveld
refinements using \textit{FullProf Suite}. The vertical marks indicate the position of Bragg reflections, and the bottom curves show the difference between the observed and calculated
intensities. The * in (b) marks the location of the (0, 0, 3/2) reflection.}
\end{figure}

\begin{figure}[tbp]
\linespread{1}
\par
\begin{center}
\includegraphics[width=\columnwidth]{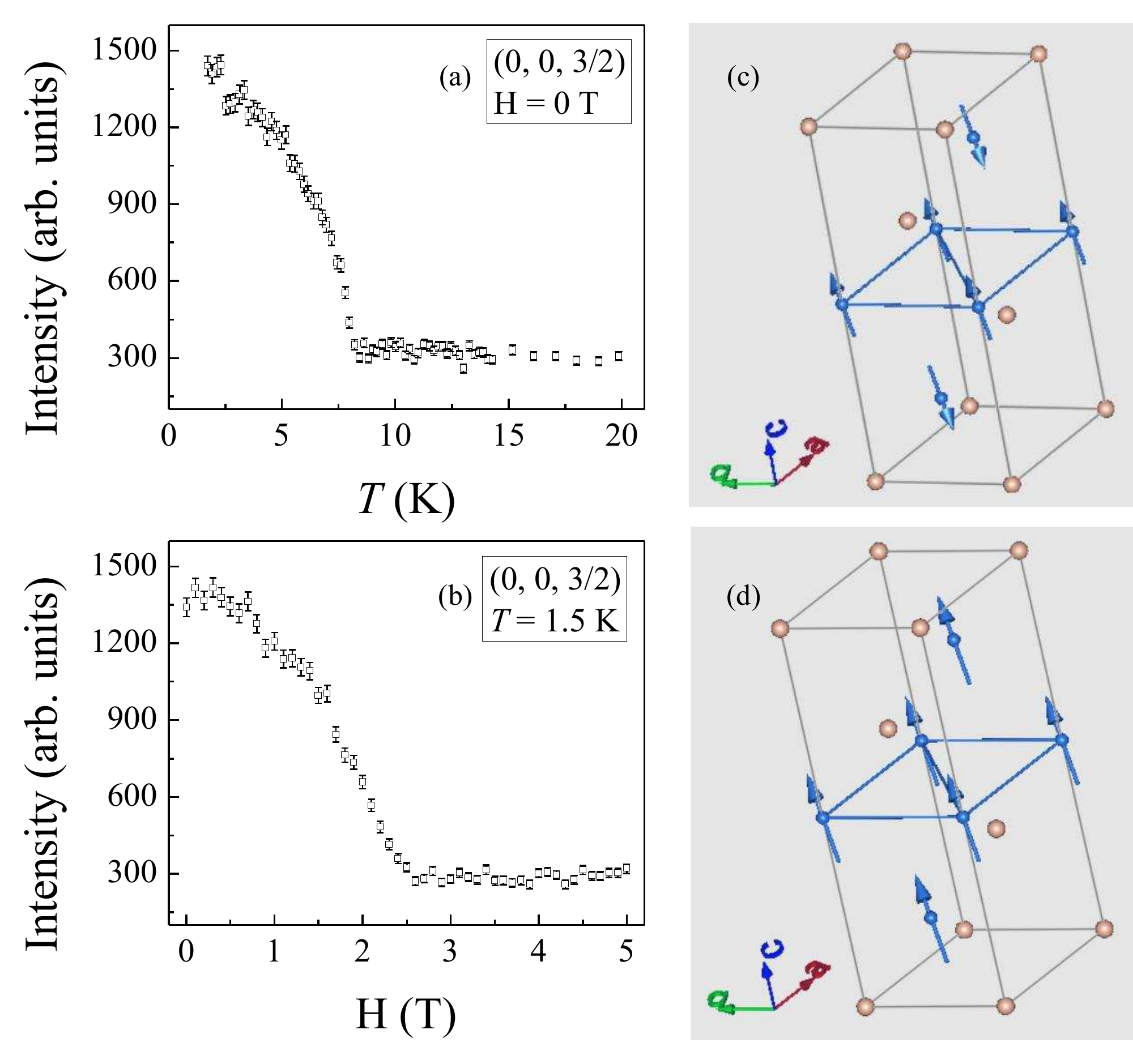}
\end{center}
\par
\caption{(color online) For YCr(BO$_3$)$_2$, (a) the temperature dependence and (b) the field dependence of the (0,0,3/2) magnetic Bragg reflection, and the magnetic ground state at (c) H = 0 T and (d) H = 5.0 T. }
\end{figure}

\begin{figure}[tbp]
\linespread{1}
\par
\begin{center}
\includegraphics[width=\columnwidth]{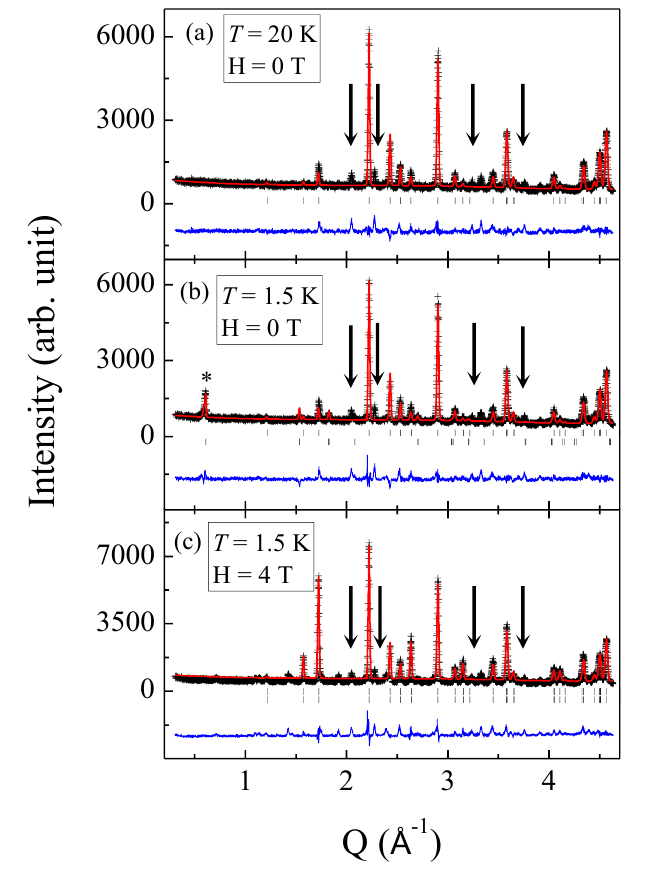}
\end{center}
\par
\caption{(color online) The neutron diffraction patterns for polycrystalline 
HoCr(BO$_3$)$_2$ (crosses) at (a) $T$ = 20 K and H = 0 T, (b) $T$ = 1.5 K and H = 0 T,
and (c) $T$ = 1.5 K and H = 4.0 T using a neutron wavelength of 2.413 {\AA}. The solid curves are the best fits from the Rietveld
refinements using \textit{FullProf Suite}. The vertical marks indicate the position of Bragg reflections, and the bottom curves show the difference between the observed and calculated
intensities. The * in (b) marks the location of the (0, 0, 3/2) reflection. The arrows denote HoBO$_3$ impurity peaks.}
\end{figure}

\begin{figure}[tbp]
\linespread{1}
\par
\begin{center}
\includegraphics[width=\columnwidth]{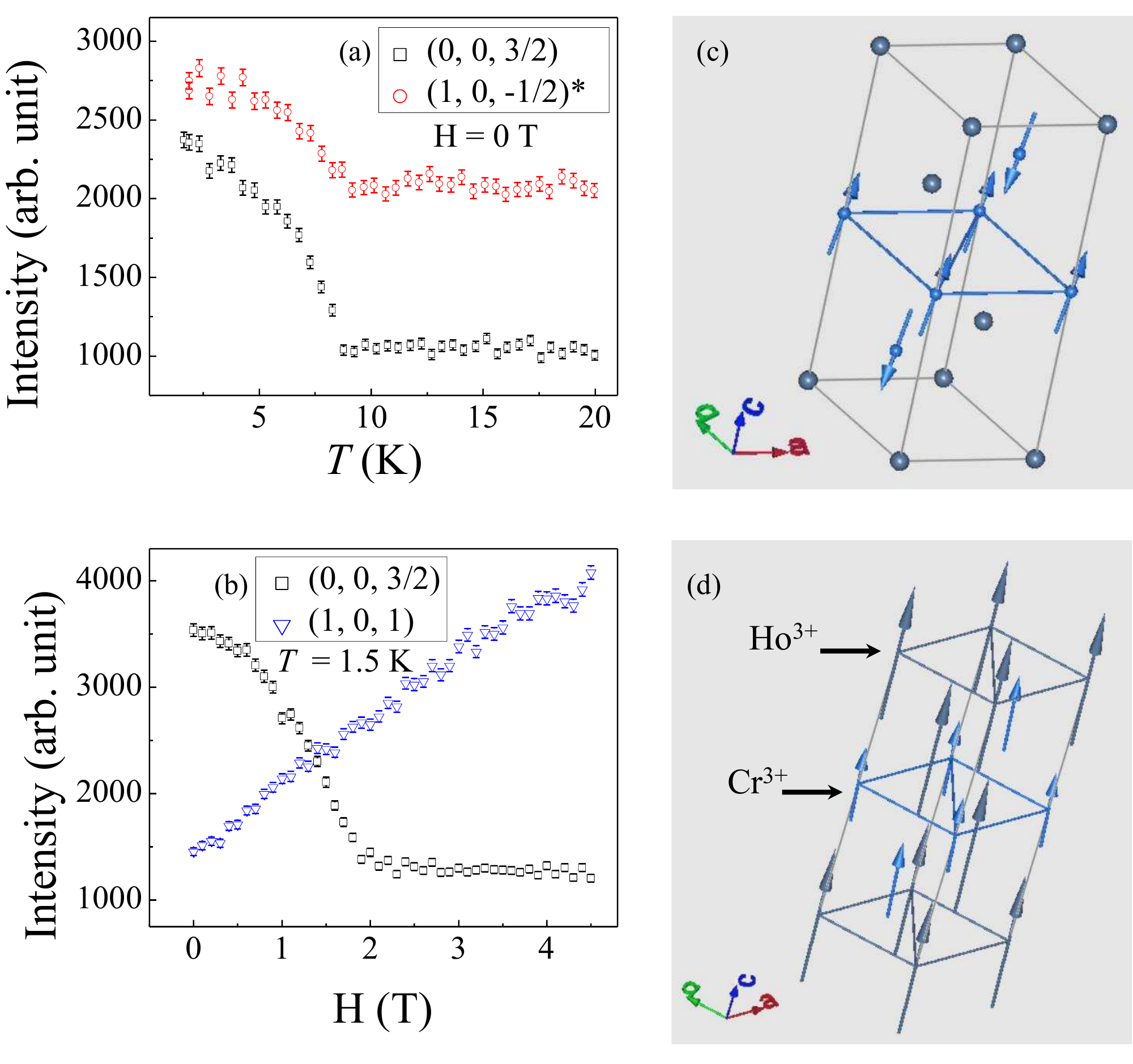}
\end{center}
\par
\caption{(color online) For HoCr(BO$_3$)$_2$, (a) the temperature dependence and (b) the field dependence of certain magnetic and lattice Bragg reflections, and the magnetic ground state at (c) H = 0 T and (d) H = 4.0 T. *Note that the reflection marked (1, 0, -1/2) in (a) also includes intensity from the (1, -1, 1/2) and (0, 1, 1/2) reflections. }
\end{figure}

\begin{figure}[tbp]
\linespread{1}
\par
\begin{center}
\includegraphics[width=\columnwidth]{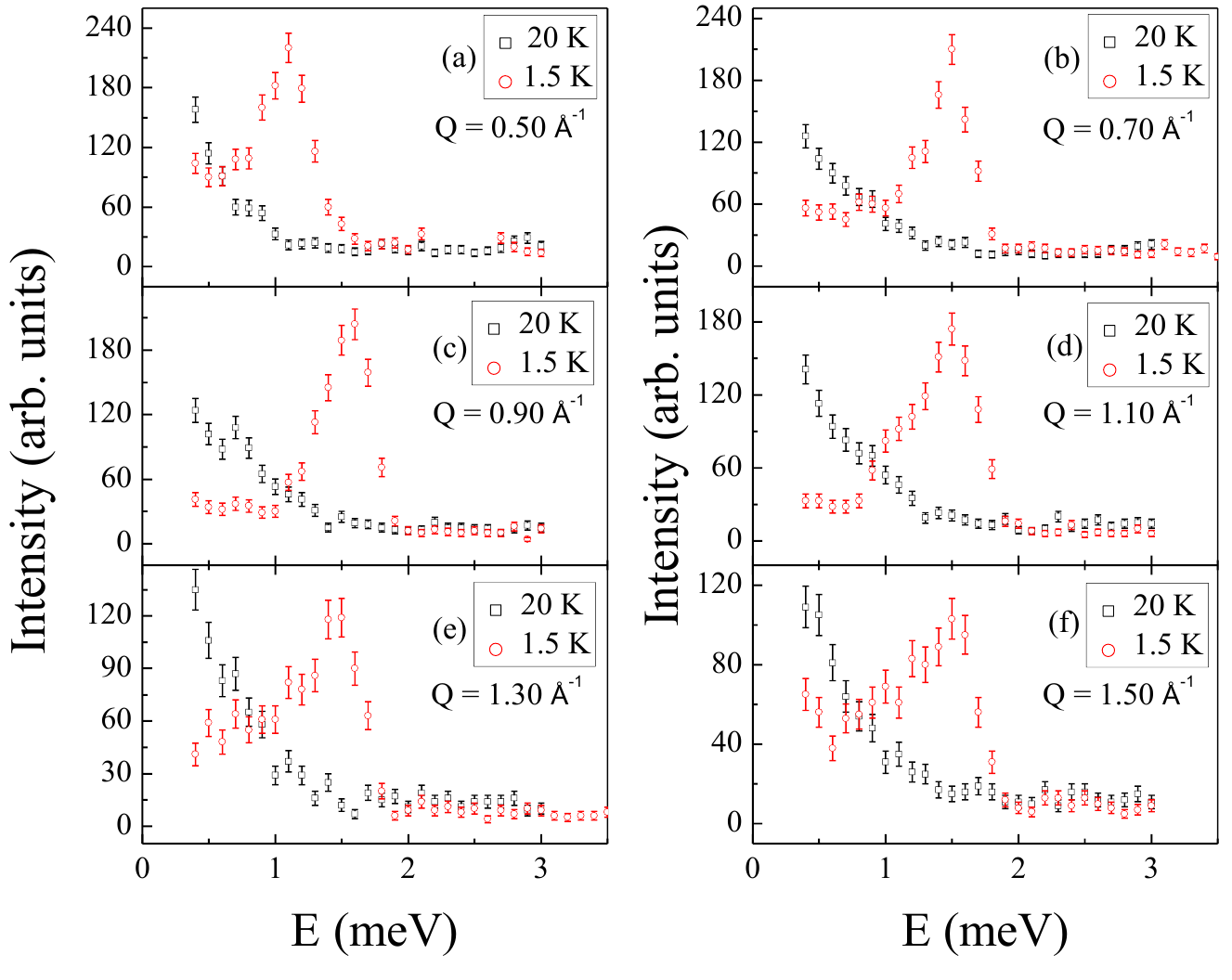}
\end{center}
\par
\caption{(color online) Inelastic neutron scattering profile above (black squares) and below (red circles) the transition temperature with E$_{F}$ = 5.0 meV centered at (a) Q = 0.50 \AA$^{-1}$, (b) Q = 0.70 \AA$^{-1}$, (c) Q = 0.90 \AA$^{-1}$, (d) Q = 1.10 \AA$^{-1}$, (e) Q = 1.30 \AA$^{-1}$, and (f) Q = 1.50 \AA$^{-1}$. Note that a spurion centered around 2.3 meV has been removed from (a).}
\end{figure}

\begin{figure*}[tbp]
\linespread{1}
\par
\begin{center}
\includegraphics[width=7 in]{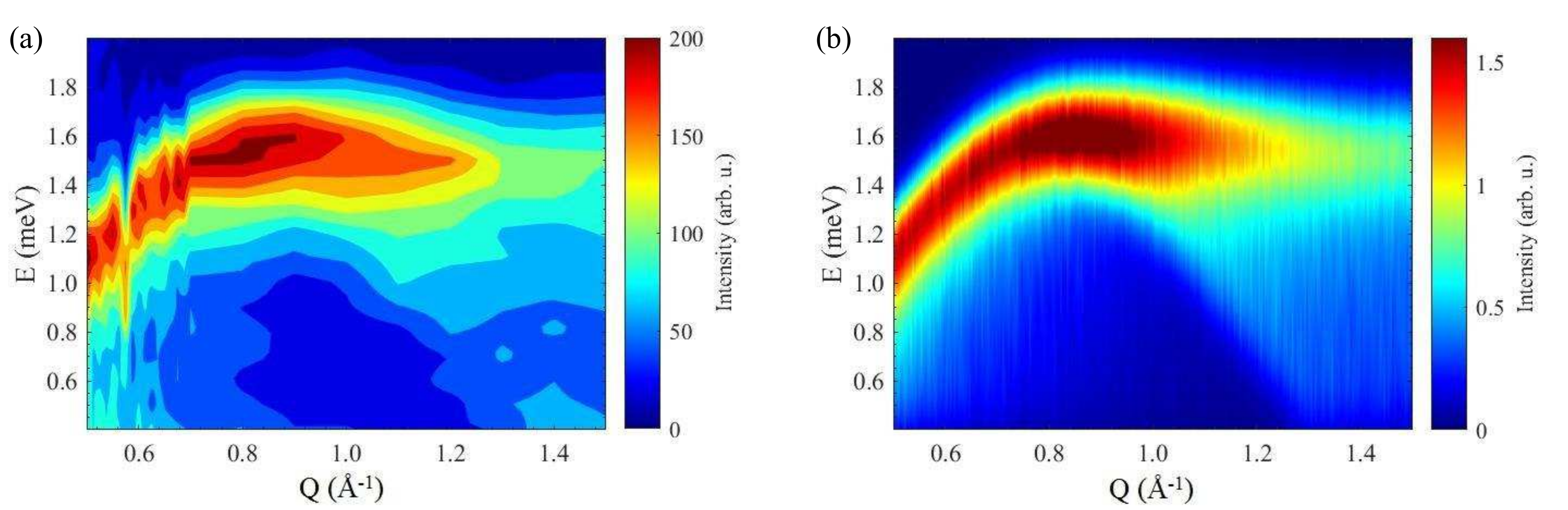}
\end{center}
\par
\caption{(color online) The (a) measured and (b) calculated powder-averaged spin wave dispersion for YCr(BO$_3$)$_2$ measured at $T$ = 1.5 K. }
\end{figure*}

\begin{figure}[tbp]
\linespread{1}
\par
\begin{center}
\includegraphics[width=2.8 in]{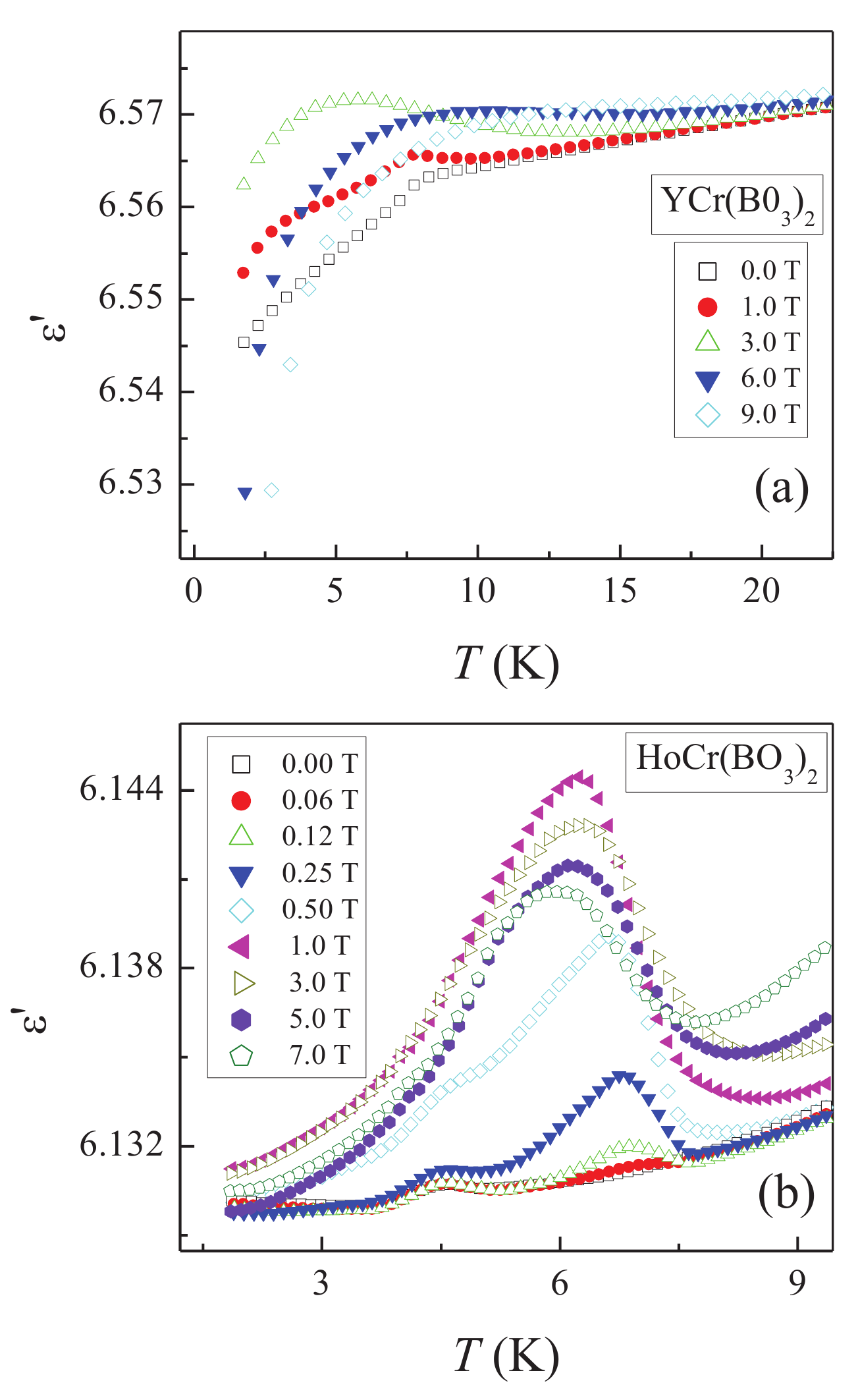}
\end{center}
\par
\caption{(color online) The temperature dependence of the dielectric constant measured at 20 kHz under applied fields for (a) YCr(BO$_3$)$_2$ and (b) HoCr(BO$_3$)$_2$. }
\end{figure}

\begin{figure}[tbp]
\linespread{1}
\par
\begin{center}
\includegraphics[width=2.8 in]{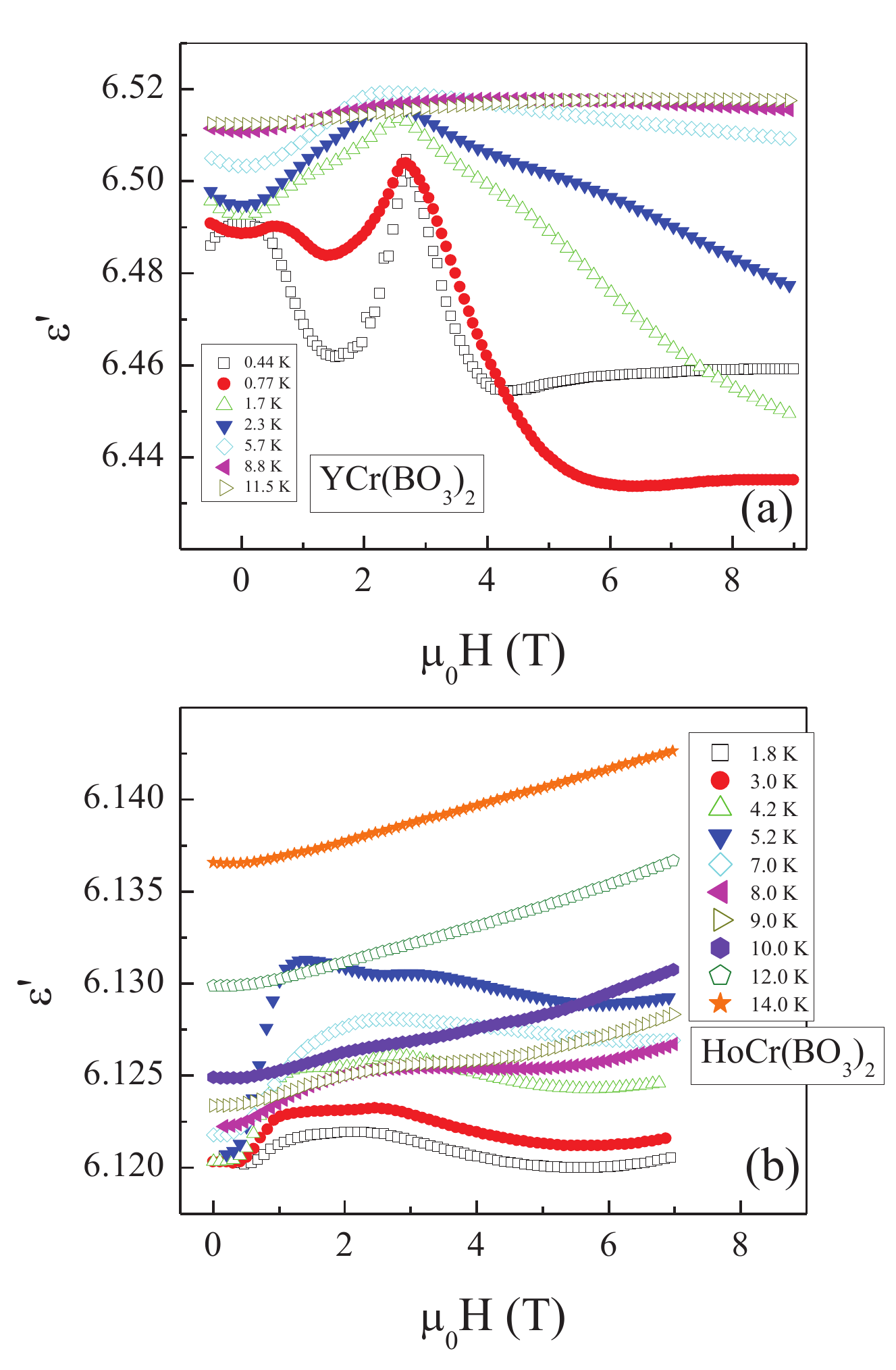}
\end{center}
\par
\caption{(color online) The field dependence of the dielectric constant measured at 20 kHz at varying temperatures for (a) YCr(BO$_3$)$_2$ and (b) HoCr(BO$_3$)$_2$.}
\end{figure}

\begin{figure}[tbp]
\linespread{1}
\par
\begin{center}
\includegraphics[width=\columnwidth]{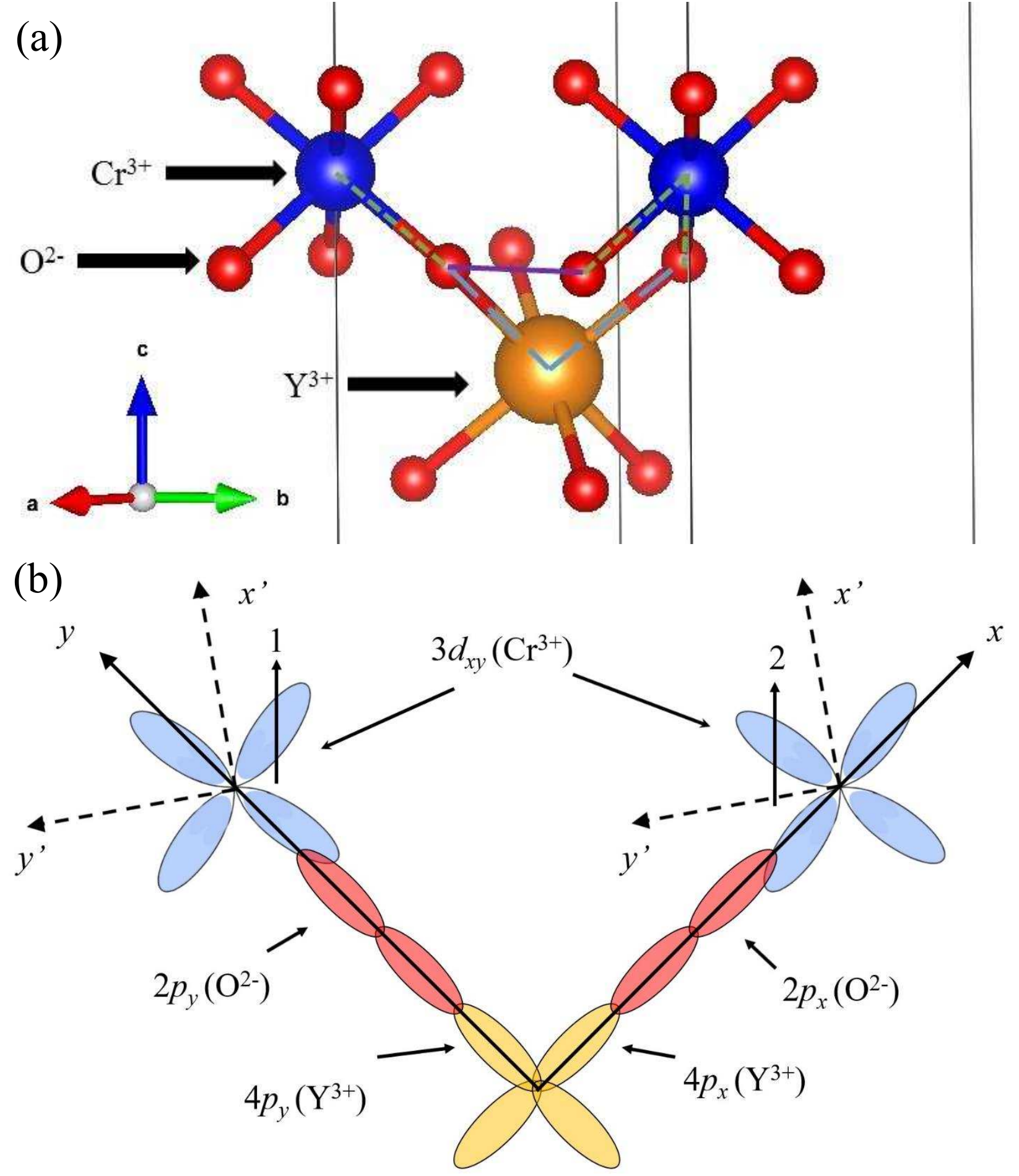}
\end{center}
\par
\caption{(color online) (a) The lattice view of the Cr$^{3+}$-O$^{2-}$-O$^{2-}$-Cr$^{3+}$ and the Cr$^{3+}$-O$^{2-}$-Y$^{3+}$-O$^{2-}$-Cr$^{3+}$ superexchange paths; (b) the orbital configurations related to the Cr$^{3+}$-O$^{2-}$-Y$^{3+}$-O$^{2-}$-Cr$^{3+}$ superexchange path. The Cr$^{3+}$ ions' frame of reference is denoted with primes. The angle between the unprimed and primed axes is 55$^\circ$.}
\end{figure}

\section{INTRODUCTION}

For many years, triangular lattice antiferromagnets (TLAFs) have been studied because of their great potential to exhibit various intriguing magnetic properties related to strong geometrical frustration.\cite{Ramirez1, Collins1, Balents1} Recently, studies of TLAFs mainly focus on four central themes: (i) quantum spin liquid (QSL) states.\cite{Anderson1, Shimizu1, Itou1, Khuntia1, Zhou1} For example, this exotic state is realized in YbMgGaO$_4$\cite{Li1, Paddison1} with an effective spin-1/2 Yb$^{3+}$ triangular lattice in which the spin anisotropy and next-nearest neighbor (NNN) interactions play an important role. In another TLAF Ba$_3$NiSb$_2$O$_9$\cite{Cheng1, Quilliam1} with Ni$^{2+}$ (S = 1), the high pressure modified phase exhibits a QSL state by suppressing the interlayer interactions. (ii) Exotic disordered states.\cite{Nakatsuji1, Honecker1, Hauke1, Yoshida1, Rachel1} One excellent example is the TLAF NiGa$_2$S$_4$\cite{Stock1} which exhibits no long range ordering. Several scenarios such as nematic or quadrupolar ground states, Kosterlitz-Thouless transitions, and Z$_2$-vortex transitions have been proposed for this disordered state.\cite{Zhao1, Yaouanc1} (iii) The coplanar 120 degree state and the related field induced spin state transitions.\cite{Ono1, Kimura1, Tocchio1, Takatsu1, Luo1} For instance, Ba$_3$CoSb$_2$O$_9$\cite{Shirata1, HDZhou1, Ma1} with an equilateral Co$^{2+}$ (effective spin-1/2) triangular lattice was found to exhibit a 120 degree ordered state at zero field and an up up down (uud) phase under a field applied in the $ab$-plane. This unique field induced state is characterized by a magnetization plateau at a value equal to 1/3 of the saturation magnetization M$_s$ and stabilized by quantum spin fluctuations. (iv) Multiferroicity.\cite{Ikeda1, Ikeda2, Elkhouni1, Lee1, Ribeiro1} In several TLAF systems, the ferroelectricity and magnetic ordering are strongly coupled. For example, the TLAF CuCrO$_2$\cite{Seki1, Sakhratov1} exhibits a spin-driven ferroelectric transition which is dependent upon the system's spiral spin structure. The TLAF RbFe(MoO$_4$)$_2$\cite{Inami1, Prozorova1, Svistov1, Kenzelmann1} shows multifeorroic properties in the 120 degree ordered state at zero field, but the ferroelectricity disappears in the field induced uud phase. The TLAF Ba$_3$ANb$_2$O$_9$ (A = Mn)\cite{Lee3} also exhibits multiferroicity exclusively in the 120 degree ordered state, while in the TLAFs Ba$_3$ANb$_2$O$_9$ (A = Co and Ni),\cite{Hwang1, Lee2} the multiferroicity not only appears at zero field but also survives in the field induced phases.

Meanwhile, multiferroic properties are also commonly associated with materials with layered structures which can greatly affect their properties, such as magnetoeletric layered perovskites.\cite{Khomskii1, Eerenstein1, Cheong1} In particular, for layered materials with more than one magnetically active ion, the exchange interactions between the different magnetic ions on different layers possibly induce strong magnetoelectric (ME) behaviors. For example,  SrNdFeO$_4$\cite{Hwang2, Oyama1} and NdCrTiO$_5$\cite{Hwang3, Greenblatt1, Saha1} are both layered perovskites with two different active magnetic ions on adjacent layers, and they both exhibit ME behaviors dependent upon the magnetic orderings or spin flop transitions. The possibility that NdCrTiO$_5$ is multiferroic is still under investigation. \cite{Tokunaga1} Therefore, from a materials engineering perspective, a material with a layered structure, two different magnetic ions, and a magnetic triangular lattice may lead to strong ME or multiferroic properties.

In search of such a material, we chose the system HoCr(BO$_3$)$_2$ to investigate. The studies of HoCr(BO$_3$)$_2$ have shown that it crystallizes in the rhombohedral space group $R\bar{3}$ and has a dolomite-type structure with a small amount of anti-site disorder between the Ho and Cr sites. \cite{Doi1} As shown in Fig. 1(a), both the Ho and Cr ions occupy octahedral sites which form a three dimensional network by sharing corner-oxygen ions. The BO$_3$ triangles also share the octahedra's oxygen atoms. Moreover, both the Ho$^{3+}$ and the Cr$^{3+}$ magnetic ions form a triangular lattice in the $ab$ plane as shown in Fig. 1(b). The intraplanar distance between the Ho/Cr ions is 4.76 {\AA}, and the interplanar distance between the Ho/Cr ions is 5.86 {\AA}. HoCr(BO$_3$)$_2$ also shows a magnetic transition around $T_N$ $\sim$ 8 K. \cite{Doi1} Therefore, HoCr(BO$_3$)$_2$ meets our requirements listed above for studies on ME behaviors. 

So far, no detailed studies on the magnetic and electric properties of this interesting system have been performed. Another advantage here is that we can replace the Ho$^{3+}$ ions with non-magnetic Y$^{3+}$ ions while retaining the same structure. Thus the comparison between the Y-compound with one magnetic ion and the Ho-compound with two magnetic ions will help us to better understand how exactly the extra magnetic ion affects the system. With this motivation, we studied the $R$Cr(BO$_3$)$_2$ ($R$ = Y and Ho) system with various experimental techniques including ac and dc susceptibility, dc magnetization, specific heat, elastic and inelastic neutron scattering, and dielectric constant measurements in order to characterize the magnetic ground states and investigate possible multiferroic properties of the system.

\section{EXPERIMENTAL METHODS}

Polycrystalline samples of $R$Cr(BO$_3$)$_2$ ($R$ = Ho and Y) were synthesized by solid state
reactions. The stoichiometric mixture of Ho$_2$O$_3$/Y$_2$O$_3$, Cr$_2$O$_3$, and 
B$_2$O$_3$ were ground together and pressed into 6-mm-diameter 60-mm rods under 400 atm
hydrostatic pressure to form rods of RCr(B$_{1.15}$O$_3$)$_2$ and then calcined in Argon at
1100 $^{\circ}$C three times: once for 12 hours and twice for 36 hours each, adding an extra
10\% of Cr$_2$O$_3$ by weight before each 36 hour annealing. 

The zero field cooling dc magnetic-susceptibility
measurements were performed using a Quantum Design superconducting interference device
(SQUID) magnetometer. The AC susceptibility was measured with a homemade setup.\cite{ZLD1} The capacitance was
measured on thin-plate polycrystalline samples with an Andeen-Hagerling AH-2700A commercial
capacitance bridge using a frequency of 20 kHz which was analyzed to obtain the dielectric constant data by approximating
the sample as an infinite parallel capacitor. The pyroelectric current was measured using a Keithley 6517A electrometer during warming after the sample was cooled in an electric field from above $T_{N}$. The specific heat measurements were performed
on a Quantum Design Physical Property Measurement System (PPMS). 

Elastic neutron scattering
measurements were performed at the Neutron Powder Diffractometer (HB-2A) using a wavelength
of either 1.5405 or 2.413 {\AA} to probe the lattice and magnetic Bragg peaks. The powder samples were pressed into rods in order to maintain the powder average in both zero field and applied fields. The inelastic neutron scattering
measurements were performed at the Cold Neutron Triple-Axis Spectrometer (CTAX) using a fixed
final energy of E$_{F}$ = 5.0 meV in order to investigate a reasonable range of energy
transfers and characterize the overall spectrum. Both instruments are located at the High
Flux Isotope Reactor (HFIR) in Oak Ridge National Laboratory (ORNL). The neutron scattering
diffraction patterns were refined by using the software packages $FullProf Suite$ and SARA$h$.\cite{FullProf, Willis1}

\section{EXPERIMENTAL RESULTS}

\begin{table*}[tbp]
\par
\caption{Magnetic moments for YCr(BO$_3$)$_2$ at $T$ = 1.5 K determined from refined neutron diffraction measurements for (a) H = 0 T and (b) H = 5.0 T.}
\label{table1}
\setlength{\tabcolsep}{0.51cm}
\begin{tabular}{ccccccccc}
\hline \hline
YCr(BO$_3$)$_2$ & Atom & x & y & z & Mx & My & Mz & M \\ \hline
\multirow{3}{*}{\begin{tabular}[c]{@{}c@{}}(a)\\ $T$ = 1.5 K\\ H = 0 T\end{tabular}} & Cr1 & 0 & 0 & 0 & 2.11(17) & 2.11(17) & 1.29(5) & 2.47(24) \\
 & Cr2 & 2/3 & 1/3 & 1/3 & -2.11(17) & -2.11(17) & -1.29(5) & 2.47(24) \\
 & Cr3 & 1/3 & 2/3 & 2/3 & -2.11(17) & -2.11(17) & -1.29(5) & 2.47(24) \\ \hline
\multirow{3}{*}{\begin{tabular}[c]{@{}c@{}}(b)\\ $T$ = 1.5 K\\ H = 5.0 T\end{tabular}} & Cr1 & 0 & 0 & 0 & 2.41(17) & 2.41(17) & 2.06(15) & 3.17(28) \\
 & Cr2 & 2/3 & 1/3 & 1/3 & 2.41(15) & 2.41(15) & 2.06(15) & 3.17(28) \\
 & Cr3 & 1/3 & 2/3 & 2/3 & 2.41(15) & 2.41(15) & 2.06(15) & 3.17(28) \\ \hline \hline
\end{tabular}
\end{table*}

\begin{table*}[tbp]
\center
\par
\caption{Magnetic moments for HoCr(BO$_3$)$_2$ at $T$ = 1.5 K determined from refined neutron diffraction measurements for (a) H = 0 T and (b) H = 4.0 T.}
\label{table2}
\setlength{\tabcolsep}{0.51cm}
\begin{tabular}{ccccccccc}
\hline \hline
HoCr(BO$_3$)$_2$ & Atom & x & y & z & Mx & My & Mz & M \\ \hline
\multirow{6}{*}{\begin{tabular}[c]{@{}c@{}}(a)\\ $T$ = 1.5 K\\ H = 0 T\end{tabular}} & Cr1 & 0 & 0 & 0 & 2.44(9) & 2.44(9) & 1.87(4) & 3.07(13) \\
 & Cr2 & 2/3 & 1/3 & 1/3 & -2.44(9) & -2.44(9) & -1.87(4) & 3.07(13) \\
 & Cr3 & 1/3 & 2/3 & 2/3 & -2.44(9) & -2.44(9) & -1.87(4) & 3.07(13) \\
 & Ho1 & 0 & 0 & 1/2 & 0.0 & 0.0 & 0.0 & 0.0 \\
 & Ho2 & 2/3 & 1/3 & 5/6 & 0.0 & 0.0 & 0.0 & 0.0 \\
 & Ho3 & 1/3 & 2/3 & 1/6 & 0.0 & 0.0 & 0.0 & 0.0 \\ \hline
\multirow{6}{*}{\begin{tabular}[c]{@{}c@{}}(b)\\ $T$ = 1.5 K\\ H = 4.0 T\end{tabular}} & Cr1 & 0 & 0 & 0 & 1.19(11) & 1.19(11) & 3.25(31) & 3.46(35) \\
 & Cr2 & 2/3 & 1/3 & 1/3 & 1.19(11) & 1.19(11) & 3.25(31) & 3.46(35) \\
 & Cr3 & 1/3 & 2/3 & 2/3 & 1.19(11) & 1.19(11) & 3.25(31) & 3.46(35) \\
 & Ho1 & 0 & 0 & 1/2 & 1.10(8) & 1.10(8) & 6.80(25) & 6.89(27) \\
 & Ho2 & 2/3 & 1/3 & 5/6 & 1.10(8) & 1.10(8) & 6.80(25) & 6.89(27) \\
 & Ho3 & 1/3 & 2/3 & 1/6 & 1.10(8) & 1.10(8) & 6.80(25) & 6.89(27) \\ \hline \hline
\end{tabular}
\end{table*}

\subsection{ac/dc Susceptibility and Specific Heat}

In Fig. 2(a), we show the temperature dependence of the dc magnetic susceptibility measured between H = 0.1 and 3.0 T for YCr(BO$_3$)$_2$. At H = 0.1 T, there is a sharp peak at $T_{N}$ = 8 K representing a magnetic transition. As the applied field is increased, $T_{N}$ decreases and the peak begins to broaden. This temperature dependence of T$_{N}$ suggests that the transition is antiferromagnetic (AFM). At H = 3.0 T, T$_{N}$ is no longer visible down to 2 K. The Curie-Weiss analysis (not shown here) of the 1/$\chi$ data above 100 K yields an effective magnetic moment of $\mu_{eff}$ = 3.85 $\mu_{B}$ and a Curie temperature of $\theta_{CW}$ = -1.86 K which are in good agreement with the previously reported values.\cite{Doi1} While $\mu_{eff}$ matches what one would expect for Cr$^{3+}$ (S = 3/2) ions, the dc magnetization measured at 2 K saturates around 2 $\mu_{B}$, smaller than the expected value of 3 $\mu_{B}$ (Fig. 2(c)). Currently, it seems that this decreased magnetic moment is likely affected by the $R$/Cr anti-site disorder which has been characterized by Doi et al.\cite{Doi1} We estimate the anti-site disorder in our sample to be $\sim$3\% from our elastic neutron scattering Rietveld refinements (as discussed below). There is also a notable slope change in the magnetization data measured at 2 K around H = 2.5 T. Moreover, Fig. 2(e) shows a kink around H = 2.5 T in the ac magnetic susceptibility data suggesting a possible phase transition near this critical value, H$_{C}$. 

Similarly, in Fig. 2(b) the temperature dependence of the dc magnetic susceptibility measured between H = 0.1 and 3.0 T for HoCr(BO$_3$)$_2$ also shows an AFM transition at T$_{N}$ = 9 K which decreases with applied field and vanishes at H = 3.0 T. The Curie-Weiss analysis of the 1/$\chi$ data above 100 K resulted in an effective magnetic moment of $\mu_{eff}$ = 10.99 $\mu_{B}$ and a Curie temperature of $\theta_{CW}$ = -15.1 K which are again in good agreement with the previously reported values.\cite{Doi1} The dc magnetization measured at 2 K shown in Fig. 2(d) saturates around 9 $\mu_{B}$, and there is a slope change which appears around H = 2.0 T. The field dependence of the ac susceptibility measured at 0.3 K shown in Fig. 2(f) also reveals a clear feature near H = 2.0 T. We again propose that these results indicate a critical field around H = 2.0 T for the system. 

The temperature dependence of the zero-field cooled specific heat measurements is plotted in Fig. 2(g-h). The data shows sharp $\lambda$-type anomalies at the same temperatures that the dc susceptibility measurements showed cusps at T$_{N}$ = 8 K and  9 K for the Y-compound and the Ho-compound, respectively. This provides further evidence that both materials undergo an AFM transition at these temperatures.

\subsection{Neutron Diffraction}

Fig. 3 shows the neutron powder diffraction (NPD) pattern measured at room temperature for YCr(BO$_3$)$_2$ and 20 K for HoCr(BO$_3$)$_2$, respectively. Both patterns were measured with a wavelength of 1.5405 {\AA} in order to study the lattice information. The refinements yielded lattice parameters $a$ = 4.76422(6) {\AA} and $c$ = 15.51574(28) {\AA} for the Y-compound and $a$ = 4.76002(6) {\AA} and $c$ = 15.49239(32) {\AA} for the Ho-compound. The refinements also show approximately 3\% site disorder between the Y/Ho and Cr sites for both samples. Moreover, a small amount of HoBO$_3$ impurity ($<$ 5\%) was observed in the Ho-compound. Both samples were also measured above ($T$ $\sim$ 20 K) and below ($T$ $\sim$ 2 K) $T_{N}$ as well as with applied fields up to H = 5.0 T at $T$ = 1.5 K using a longer wavelength of 2.413 {\AA} in order to study the magnetic structure information. The samples were prepared as pressed rods to maintain the powder average under applied fields. The diffraction patterns and refinements for the Y-compound and the Ho-compound can be seen in Fig. 4(a-c) and Fig. 6(a-c), respectively. 

For YCr(BO$_3$)$_2$ which contains only a single magnetic ion (Cr$^{3+}$), the NPD pattern clearly shows extra Bragg peaks at 2 K under zero field (Fig. 4(b)). These lattice forbidden reflections strongly suggest an AFM spin structure and they can be described by a propagation vector k = (0, 0, 3/2).  The magnetic structure resulting from the diffraction pattern refinement is shown in Fig. 5(c). The Cr$^{3+}$ spins form a ferromagnetic configuration in the $ab$-plane but tilt away from the $ab$ plane with a canting angle of 31.5$\degree$. Between the layers, the Cr$^{3+}$ planes align antiferromagnetically.

The intensity of the (0, 0, 3/2) magnetic Bragg reflection was investigated as a function of both temperature and magnetic field as shown in Fig. 5(a-b). The reflection not only disappears above $T_{N}$ but also is suppressed by a critical magnetic field H$_{C}$ = 2.5 T which agrees with our previous dc and ac susceptibility measurements. The NPD pattern measured at 2 K under H = 5.0 T (Fig. 4(c)), larger than H$_{C}$, confirms that the material now adopts a ferromagnetic (FM) ground state which is supported by the observed magnetic Bragg peaks which are at the same positions as the lattice Bragg peaks. This suggests a new propagation vector, k$_{FM}$ = (0 0 0).  As shown in Fig. 5(d), the refinement of the 5.0 T NPD pattern shows that in this FM state, the Cr$^{3+}$ spins are aligned in the $ab$-plane with a canting angle of 40.6$\degree$ away from the $ab$-plane. Therefore, a magnetic field above H$_{C}$ flips the AFM arrangements of spins between the layers along the $c$ axis at zero field to ferromagnetic alignments. As shown in Table I, the total magnetic moment for both AFM and FM magnetic ground states is $\mu_{Cr}$ $\sim$ 2.5 $\mu_B$ which is close to the theoretical value for Cr$^{3+}$ ions as well as the effective magnetic moment derived from our 1/$\chi$ data. 

While HoCr(BO$_3$)$_2$ contains two magnetic ions (Cr$^{3+}$/Ho$^{3+}$), its magnetic structure at zero field is very similar to that of the Y-compound's. Below $T_N$ with no applied field, the system is again described by a propagation vector of k = (0, 0, 3/2) (Fig. 6(b)). Analogously to the Y-compound, the refinement shows that for HoCr(BO$_3$)$_2$, the Cr$^{3+}$ spins arrange ferromagnetically with a canting angle of 37.5$\degree$ away from the $ab$-plane while the Cr$^{3+}$ layers align antiferromagnetically (Fig. 7(c)). This canting angle is slightly larger than that of the Y-compound. Here, no evidence was observed to support the magnetic ordering of the Ho$^{3+}$ spins down to 2 K at zero field. The total refined magnetic moment of $\mu_{Cr}$ = 3.07(13) $\mu_B$ as shown in Table II(a) supports this conclusion as the moment size is close to the theoretical value for Cr$^{3+}$ ions. 

Major magnetic Bragg reflections of HoCr(BO$_3$)$_2$ were also investigated. As shown in Fig. 7(a), the intensities of the (0, 0, 3/2) and the (1, 0, -1/2), (1, -1, 1/2), and (0, 1, 1/2) reflections are suppressed above $T_{N}$ = 9 K. Meanwhile at 2 K, the intensity of (0, 0, 3/2) is suppressed above H$_{C}$ = 2.0 T while the intensity of the (1, 0, 1) reflection increases linearly with increasing field. These critical values  agree with our previous susceptibility measurements. 

From Fig. 6(c) we can see that the lattice Bragg peaks and magnetic Bragg peaks align exactly for HoCr(BO$_3$)$_2$ at 2 K and under 4.0 T. This suggests that the system enters a FM ground state above H$_{C}$ similar to the Y-compound. However, the refinement of this data reveals that both the Cr$^{3+}$ and the Ho$^{3+}$ spins order now. The obtained spin structure at 2 K and under 4.0 T is shown in Fig. 7(d) in which both the Ho$^{3+}$ and  Cr$^{3+}$ spins are arranged ferromagnetically in the $ab$ plane, but the Cr$^{3+}$ spins have a canting angle of 69.8$\degree$ away from the $ab$-plane while the Ho$^{3+}$ ions have a canting angle of 80.8$\degree$ away from the $ab$-plane. The total refined magnetic moment of 10.35(44) $\mu_B$ calculated from Table II(b) also supports the fact that now both the Cr$^{3+}$ and the Ho$^{3+}$ spins order and contribute to the value of the magnetic moment. This value also matches closely with the saturation value determined from the magnetization curve as well as the effective magnetic moment derived from our 1/$\chi$ data. 

By analyzing the magnetic structure information obtained from the neutron diffraction data combined with the previously determined structural information, we were able to obtain two possible magnetic space groups, $R_{\textit{I}}\bar{3}$ and $P_{\textit{S}}\bar{1}$, using the Bilbao Crystallographic Server \cite{BCS1, BCS2, BCS3, k-SUBGROUPSMAG}. While the $P_{\textit{S}}\bar{1}$ magnetic space group allows for the magnetic moment to freely align along any direction,  $R_{\textit{I}}\bar{3}$ completely restricts the magnetic moment to the $c$-axis. Moreover, the Rietveld refinements of the system reveal that the existence of the intense (0, 0, 3/2) peak depends upon having a magnetic moment in the $ab$-plane. Therefore, our data strongly suggests that the system is best described by the $P_{\textit{S}}\bar{1}$ magnetic space group.

\subsection{Inelastic Neutron}

Fig. 8 shows the inelastic neutron scattering profiles measured at 20 K and 1.5 K with various momentum transfers (Q) ranging from 0.5  {\AA}$^{-1}$ to 1.5  {\AA}$^{-1}$. At each Q, a peak in the intensity with the energy transfer (E) between 1-2 meV is clearly observed at 1.5 K which should represent the spin wave excitation in the magnetic ordered state. This feature disappears in the 20 K data suggesting that the observed peak is from the magnetic origin.  

The spin wave excitation was analyzed to produce a spin wave spectrum within a limited E-Q space as shown in Fig. 9(a). The first branch of the spectrum is visible in our region of interest. The feature flattens out around Q = 0.8 {\AA}$^{-1}$ and peaks near 1.5 meV. From the location of the magnetic peaks in Fig. 4(b), we expect the first zone boundary of the spectrum to be centered near Q = 0.60  {\AA}$^{-1}$ and the second zone boundary to be centered near Q = 1.53  {\AA}$^{-1}$. 

In order to simulate this spin wave spectrum, we used the Matlab library SpinW to model the system.\cite{SpinW} SpinW uses classical Monte Carlo simulations as well as linear spin wave theory in order to solve the spin Hamiltonian: 
\begin{equation}
H=‎‎\sum_{i,j} S_{i}J_{ij}S_{j}
\end{equation}
where $S_{i}$ are spin vector operators and $J_{ij}$ are 3x3 matrices which describe pair coupling between spins.

For our model, we examined the nearest neighbor (NN) intralayer interaction J$_1$ as well as the next nearest neighbor (NNN) interlayer interaction J$_2$. The values for J$_1$ and J$_2$ were determined empirically by comparing the simulated spin wave spectrum against the experimental data. J$_1$ = -0.12 meV determines the general size and location of the feature, and J$_2$ = 0.014 meV determines the slope of the branch. Specifically, J$_2$ changed the initial energy value where the branch begins at Q = 0.50 {\AA}$^{-1}$ and the maximum energy value of the branch achieved near Q = 0.90 {\AA}$^{-1}$. Several values of J$_1$ and J$_2$ were tested in order to closely match the experimental results.

The final result is shown in Fig. 9(b). The simulation was constructed using a finite energy resolution consistent with the elastic line for E$_{F}$ = 5.0 meV of dE = 0.3 meV . Similar to the measured data, the simulation shows one branch which begins near 1.0-1.2 meV at Q = 0.50 {\AA}$^{-1}$, flattens around Q = 0.80 {\AA}$^{-1}$, and fades into the second zone boundary. Moreover, both the measured and the calculated data show very little intensity below 1.0 meV of transferred energy. Although we expected to find a spin gap with a magnitude close to 0.5 meV,  both the measured data and the simulation appear to be gapless. Our simulation provides a good overall agreement with the experimental data which reasonably suggests that the intralayer interaction is about one order of magnitude stronger than the interlayer interaction in YCr(BO$_3$)$_2$.

\subsection{Dielectric Constant}

Fig. 10(a) shows the temperature dependence of the dielectric constant, $\varepsilon$, for YCr(BO$_3$)$_2$. At zero field, $\varepsilon$ shows a cusp near 8 K. While this feature broadens consistently with the strength of the applied field, the amplitude and transition temperature are more complicated. The amplitude increases with increasing applied field up to the critical field H = 3.0 T at which point it begins to decrease as the applied fields get even larger. On the other hand, the transition temperature decreases with increasing applied field up to H = 3.0 T and then increases with larger applied fields. The temperature dependence of $\varepsilon$ for HoCr(BO$_3$)$_2$ exhibits a  more drastic response. In Fig. 10(b), a broad feature around $T$ = 4 K is observed at zero field. As the field increases, this shoulder feature becomes suppressed and vanishes near H = 1.0 T. Meanwhile, a sharp peak appears near $T$ = 7 K which increases in relation to the field up to H = 1.0 T at which point it begins to weaken with increasing field. Furthermore, the critical temperature associated with the sharp peak decreases with increasing field. 

Fig. 11(a) shows the magnetic field dependence of $\varepsilon$ for YCr(BO$_3$)$_2$. Below $T$ = 1.5 K, the data shows a sharp peak around H = 3.0 T as well as a clear minimum near H = 1.5 T. Above $T$ = 1.5 K, only the sharp peak near H = 3.0 T remains visible. Above $T_{N}$, this behavior disappears. The magnetic field dependence of $\varepsilon$ for HoCr(BO$_3$)$_2$ is presented in Fig. 11(b). At temperatures lower than its $T_{N}$, $\varepsilon$ increases sharply at low fields and then saturates into a broad feature around H = 1.5 T. Similar to the Y-compound, above its transition temperature, such behavior disappears.

These anomalies observed from $\varepsilon$ for both the Ho-compound and the Y-compound are all observed around their magnetic ordering temperatures or critical fields for spin state transitions; therefore, both systems exhibit some degree of magnetodielectric (MD) coupling. Furthermore, the replacement of the non-magnetic Y$^{3+}$ ion with the magnetic Ho$^{3+}$ ion affects this coupling which leads to stronger MD phenomena as revealed by the sharp peak around $T_{N}$ in the field induced $\varepsilon$ data. It is also worth noting that the Ho-compound shows little response at zero field near $T_{N}$ in stark contrast to the Y-compound. Thus the MD phenomena are likely  related to different mechanisms for each sample, such as spin-phonon coupling or magnetostriction.

While both compounds were studied via pyroelectric current measurements at different magnetic fields, no electric polarization was observed for either bulk polycrystalline sample around the transition temperatures. It is possible that a single crystal sample could produce an anisotropic polarization which is hidden by the powder averaging of the results or that the system is already ordered in an antiferroelectric state. Further experiments including Polarization vs. Electric field hysteresis measurements on single crystal samples may be necessary to elucidate the matter.

\section{DISCUSSION}

The two compounds share several characteristics of their magnetic properties. The Cr$^{3+}$ spins of both samples enter a canted AFM state below $T_N$ $\sim$ 9 K at zero field. With applied field above a critical vale of H$_{C}$ $\sim$  2.0 to 2.5 T, the antiferromagnetic arrangement of the Cr$^{3+}$ spins along the $c$-axis is flipped to become ferromagnetic for both samples. The major difference here is that for the Ho-compound, both the Ho$^{3+}$ and the Cr$^{3+}$ spins order ferromagnetically  when H $>$ H$_{C}$. This canted AFM state at zero field with spins aligning ferromagnetically in the $ab$-plane and antiferromagnetically along the $c$-axis of YCr(BO$_3$)$_2$ is consistent with the fact that its intralayer interaction is ferromagnetic and interlayer ineraction is antiferromagnetic which was revealed by the spin wave spectrum simulation.

To understand why the intralayer interaction of the Y-compound is ferromagnetic we look into the superexchange interactions involving the Cr$^{3+}$ ions. In order to qualitatively discuss the sign (FM or AFM) of the superexchange interactions, we turn to Kanamori theory. For a magnetic cation on an octahedral site, Kanamari has shown that the superexchange interaction via nonmagnetic anion is closely connected with the orbital states of the cation and anion.\cite{Kanamori_Theory} In YCr(BO$_3$)$_2$, two superexchange pathways for the Cr$^{3+}$ spins in the same layer are available as the CrO$_6$ octahedrons are connected by YO$_6$ octahedrons with corner sharing oxygens. As shown in Fig. 12(a), the first is Cr$^{3+}$-O$^{2-}$-O$^{2-}$-Cr$^{3+}$ and the second is Cr$^{3+}$-O$^{2-}$-Y$^{3+}$-O$^{2-}$-Cr$^{3+}$. As observed from other magnetic oxides, the Cr$^{3+}$-O$^{2-}$-O$^{2-}$-Cr$^{3+}$ pathway's superexchange interaction  is often AFM. Meanwhile, one possible situation for the Cr$^{3+}$-O$^{2-}$-Y$^{3+}$-O$^{2-}$-Cr$^{3+}$ exchange path is shown in Fig. 12(b). 

Here we consider the superexchange interaction between the spins on the $d_{xy}$ orbitals of the Cr$^{3+}$ ions. In the Cr$^{3+}$ ions' frame of reference, the $d_{xy}$ orbitals are centered 45$^\circ$ from both the $x$-axis and the $y$-axis. Through our Rietveld refinements of the neutron diffraction pattern, we determined that the Cr$^{3+}$-O$^{2-}$-Y$^{3+}$ bond angle is 123.67(9)$^\circ$ and the O$^{2-}$-Y$^{3+}$-O$^{2-}$ bond angle is 88.79(9)$^\circ$, very close to 90$^\circ$; therefore, the $d_{xy}$ orbitals are centered $\sim$10$^\circ$ from the line where the O$^{2-}$ and the Y$^{3+}$ ions are situated which allows for the necessary hybridization to occur between the $d_{xy}$ and the $p_x$ and $p_y$ orbitals. In this configuration, the spin 1 on the left Cr$^{3+}$ ion is transferred to the molecular orbital composed of the $p_y$ orbitals of the O$^{2-}$ 2$p$ orbitals and the Y$^{3+}$ 4$p$ orbitals (the filled outermost orbitals) and the spin 2 on the right Cr$^{3+}$ ion is transferred to the molecular orbital composed of the $p_x$ orbitals of the O$^{2-}$ and Y$^{3+}$ ions. Due to Hund's rules, these two spins on the $p_y$ and $p_x$ orbitals in the Y$^{3+}$ ions have to be parallel. Then, after these two spins are transferred back to the Cr$^{3+}$ ion, a FM superexchange interaction is built. For YCr(BO$_3$)$_2$, it is reasonable to assume that this FM interaction overcomes the AFM interaction leading to the FM spin arrangements in $ab$-plane. 

In several other TLAF systems with layered perovskite structures, similar FM superexchange interactions involving 3$d$-2$p$-4$p$ (or 3$p$)-2$p$-3$d$ paths have been reported. For example, in Ba$_3$CoNb$_2$O$_9$,\cite{Yokota1} a weak AFM interaction is the result of the FM Co$^{2+}$-O$^{2-}$-Nb$^{5+}$-O$^{2-}$-Co$^{2+}$ superexchange interaction involving the Nb$^{5+}$ 4$p$ orbitals competing with the AFM Co$^{2+}$-O$^{2-}$-O$^{2-}$-Co$^{2+}$ interaction. Accordingly, this system exhibits a small saturation field and a low AFM transition temperature. In another triangular lattice magnet, AAg$_2$M(VO$_4$)$_2$ (A=Ba, Sr; M=Co, Ni),\cite{Moller1} the FM  Co$^{2+}$-O$^{2-}$-V$^{5+}$-O$^{2-}$-Co$^{2+}$ interaction involving the V$^{5+}$ 3$p$ orbitals is stronger than the AFM Co$^{2+}$-O$^{2-}$-O$^{2-}$-Co$^{2+}$ interaction resulting in a FM transition. One important note here for YCr(BO$_3$)$_2$ is that although the FM interaction overcomes the AFM interaction in the $ab$-plane, the AFM interlayer interaction still leads to an AFM arrangement of spins along the $c$-axis to stabilize the canted AFM spin structure. This interlayer interaction is weaker, but it plays an important role in defining the magnetic ground state.

Both YCr(BO$_3$)$_2$ and HoCr(BO$_3$)$_2$ show some MD behaviors. The Y-compound's dielectric constant shows a slope change around $T_{N}$ and a sharp peak around H$_{C}$; on the other hand, the Ho-compound's dielectric constant shows a strong peak around $T_{N}$ with an applied field and a broad peak around H$_{C}$. Apparently, this difference is related to the presence of the second magnetic ion, Ho$^{3+}$, in HoCr(BO$_3$)$_2$. To explain the MD anomalies, we first examined a possible linear magnetoelectric (ME) effect. For our system, both possible magnetic space groups, $R_{\textit{I}}\bar{3}$ and $P_{\textit{S}}\bar{1}$, contain an inversion center as one of their symmetry elements. Additionally, symmetry operators in both magnetic space groups do not break
time reversal symmetry. Therefore, the linear ME effect is excluded by symmetry.

We also observed that the cusp-shape of the Y-compound's MD anomaly is similar to behavior observed in the AFM EuTiO$_3$.\cite{Katsufuji1, Shvartsman1} In EuTiO$_3$, the pair correlation of the Eu spins to a soft-phonon mode containing Eu-O stretching motions was ascribed to the MD anomaly. Such spin-phonon coupling was also attributed to the MD anomaly observed in a ferrimagnetic spinel Mn$_3$O$_4$.\cite{Tackett1}
On the other hand, it is also possible that the higher order, symmetry independent ME terms can be relevant as in the case for TeCuO$_3$\cite{Lawes1} and in Cr[(H$_3$N-(CH$_2$)$_2$-PO$_3$(Cl)(H$_2$O)].\cite{Nenert1, Nenert2}

Another way to understand the differences between the MD effect in both samples is to consider magnetostriction. As the order of magnitude of the effect is fairly large ($\sim$ 10$^{-3}$), we speculate that the differences in the dielectric constant data are more likely due to the magnetostriction caused by the extra exchange interaction between the Cr$^{3+}$ and Ho$^{3+}$ layers with an applied field rather than due to the change of the lattice parameters which typically produces a much smaller anomaly (10$^{-5}$ $\sim$ 10$^{-6}$). Specifically, (i) at zero field, there is no exchange interaction between the Cr$^{3+}$ and Ho$^{3+}$ spins in HoCr(BO$_3$)$_2$ since only the Cr$^{3+}$ spins order. Therefore, there is no obvious dielectric anomaly around $T_{N}$; (ii) with an applied field H $<$ H$_{C}$, the short range ordering of Ho$^{3+}$ spins could be induced which can lead to an AFM exchange interaction between the Cr$^{3+}$ and Ho$^{3+}$ layers and result in magnetostriction. Thus, a small magnetic field such as 0.25 T induces a dielectric constant peak around $T_{N}$. Moreover, with increasing field this effect is strengthened by involving more short range ordered Ho$^{3+}$ spins, and, consequently, the dielectric constant peak intensity increases; (iii) with even larger applied fields  H $>$ H$_{C}$, the Ho$^{3+}$ spins order ferromagnetically along with the Cr$^{3+}$ spins. This new spin structure possibly leads to weak magnetostriction compared to that of H $<$ H$_{C}$. Therefore, the dielectric constant peak intensity achieves the highest value with H = 1.0 T and then decreases with increasing field as soon as it exceeds H$_{C}$, such as 3.0 T. 

In any case, more studies are needed to determine the origin of the observed MD anomaly in both compounds. Experimental probes such as infrared and Raman spectroscopy could reveal possible spin-phonon coupling. Furthermore, dielectric constant and polarization (pyroelectric current) measurements on single crystal samples can be helpful not only to identify the ME coefficients for both compounds but also to study the possible magnetostriction effect for the Ho-compound.

\section{CONCLUSION}

In summary, we report detailed experimental studies of the layered perovskites $R$Cr(BO$_3$)$_2$ ($R$ = Y and Ho) with triangular lattices, focusing on their magnetic and electric properties. We observed the presence of a canted AFM state in both samples at zero field for the Cr$^{3+}$ spins as well as a FM state while a critical field was applied. More interestingly, in comparison to the Y-compound, far different MD behaviors were observed in the Ho-compound which should be due to the interplay between the Cr and Ho magnetic layers. Our studies here demonstrate that the combination of layered structures with two different magnetic ions and triangular lattices can produce intriguing physical properties. This principle of materials engineering can help us to design and explore  more complex magnetic materials.

\begin{acknowledgments}
R. S. and H.D.Z. thank the support from NSF-DMR through Award DMR-1350002. R. S. and H.D.Z would also like to thank Dr. Stephen Nagler at ORNL for many enlightening conversations regarding this system. This research used resources at the High Flux Isotope Reactor, a DOE Office of Science User Facility operated by the Oak Ridge National Laboratory. A portion of this work was performed at the NHMFL, which is supported by National Science Foundation Cooperative Agreement No. DMR-1157490 and the State of Florida. E. S. Choi and M. Lee acknowledge the support from NSF-DMR-1309146. 
\end{acknowledgments}

\end{document}